\documentclass[conference]{IEEEtran}
\IEEEoverridecommandlockouts
\usepackage{arydshln}
\usepackage{cite}
\usepackage{amsmath,amssymb,amsfonts}
\usepackage{algorithmic}
\usepackage{graphicx}
\usepackage{textcomp}
\usepackage{subfigure}
\usepackage{xcolor}
\usepackage{marvosym}
\def\BibTeX{{\rm B\kern-.05em{\sc i\kern-.025em b}\kern-.08em
    T\kern-.1667em\lower.7ex\hbox{E}\kern-.125emX}}
\begin{document}

\title{Exploring Downvoting in Blockchain-based \\ Online Social Media Platforms
}

\author{
  \IEEEauthorblockN{
    Rui~Sun\IEEEauthorrefmark{1}\IEEEauthorrefmark{2},
    Chao~Li\IEEEauthorrefmark{1}\IEEEauthorrefmark{2},
    Jingyu~Liu\IEEEauthorrefmark{1}\IEEEauthorrefmark{2},
    Xingchen~Sun\IEEEauthorrefmark{1}\IEEEauthorrefmark{2}
  }
  \IEEEauthorblockA{
  \IEEEauthorrefmark{1}Beijing Key Laboratory of Security and Privacy in Intelligent Transportation, Beijing Jiaotong University, China \\
    \IEEEauthorrefmark{2}School of Computer and Information Technology, Beijing Jiaotong University, China \\
    \{21120483, li.chao, 22125196, 22120504\}@bjtu.edu.cn
  }
}

\maketitle

\begin{abstract}
In recent years, Blockchain-based Online Social Media (BOSM) platforms have evolved fast due to the advancement of blockchain technology.
  BOSM can effectively overcome the problems of traditional social media platforms, such as a single point of trust and insufficient incentives for users, by combining a decentralized governance structure and a cryptocurrency-based incentive model, thereby attracting a large number of users and making it a crucial component of Web3.
  BOSM allows users to downvote low-quality content and aims to decrease the visibility of low-quality content by sorting and filtering it through downvoting.
  However, this feature may be maliciously exploited by some users to undermine the fairness of the incentive, reduce the quality of highly visible content, and further reduce users' enthusiasm for content creation and the attractiveness of the platform.
  In this paper, we study and analyze the downvoting behavior using four years of data collected from Steemit, the largest BOSM platform.
  We discovered that a significant number of bot accounts were actively downvoting content. 
  In addition, we discovered that roughly 9\% of the downvoting activity might be retaliatory. 
  We did not detect any significant instances of downvoting on content for a specific topic.
  We believe that the findings in this paper will facilitate the future development of user behavior analysis and incentive pattern design in BOSM and Web3.
\end{abstract}


\section{Introduction}
Traditional Online Social Media (OSM) platforms, such as Facebook~\cite{b1} and Twitter~\cite{b2}, have two main problems.
First, these platforms are highly centralized and thus subject to a single point of trust. Users are forced to trust that the platform will not misuse or disclose their data, that their posts and comments will not be tampered with or deleted, and that their accounts will not be forcibly restricted or blocked. 
However, these situations have frequently occurred in reality~\cite{b3}, and the underlying reason is that users do not have the right to manage the platform.
Second, these platforms rely on users to create valuable content for profit, but as providers of content, users rarely receive adequate rewards~\cite{b4}. This traditional business model lacks effective incentives for users to provide high-quality content, resulting in a mismatch between efforts and rewards and, thus, decreased motivation and creativity.

In recent years, blockchain technology has facilitated the fast growth of Blockchain-based Online Social Media (BOSM) platforms.
Blockchain is a distributed ledger that manages user transactions over a peer-to-peer network, with features including tamper-resistance and decentralization~\cite{b5}.
Based on these features, blockchain can effectively protect BOSM platforms from problems such as a single point of failure and the lack of authenticity of platform content.
On this basis, blockchain allows users to participate in the management of BOSM platforms and grants them the right to manage the platform, thus overcoming the single point of trust problem.
In addition, based on the economic properties of blockchain, BOSM platforms can use cryptocurrency to incentivize users, which encourages them to produce high-quality content.
In the past few years, a number of BOSM platforms have emerged, such as Sapien\footnote{https://www.sapien.network/}, Peepeth\footnote{https://peepeth.com/welcome}, Steemit\footnote{https://steemit.com/}, Indorse\footnote{https://indorse.io/}, Hive blog\footnote{https://hive.blog/}, etc. 
All of these platforms have the feature of decentralized platform management and the combination of underlying cryptocurrency networks and social networks~\cite{b6}.

BOSM allows users to downvote low-quality content. 
The essence of downvoting in BOSM is to efficiently rank content based on quality by crowdsourcing the task of content quality screening to users and giving the content the appropriate visibility, thus increasing the visibility of high-quality content to attract users to read it and decreasing the visibility of low-quality content to filter spam.
However, some users may abuse this feature. 
For example, users may downvote substantially more frequently than normal. 
Users may downvote for reasons unrelated to content quality, such as the publisher's identity or behavior.
In addition, users may organize to downvote information associated with an opposing topic or opinion in order to restrict its visibility. 
These downvote abuses weaken the fairness of the incentives, lower the quality of highly visible content, and further reduce users' enthusiasm for content creation and the attractiveness of the platform.

This paper presents the first empirical study on downvoting in BOSM platforms utilizing four years of data collected from Steemit~\cite{b7}, the largest BOSM platform.
Specifically, we first evaluated individual downvoting behaviors to identify active downvoters and suspected bot accounts.
We discovered that a significant number of suspected bot accounts were actively downvoting content. 
Second, we examined downvoting behaviors between users.
Our findings imply that around 9\% of downvoting might be retaliatory.
Lastly, we identified and examined topics from posts that were downvoted.
According to the findings, there are no notable instances of content downvoting for a particular topic.
We believe that the findings in this paper will facilitate the future development of user behavior analysis and incentive pattern design in both BOSM and Web3.

We organize the rest of this paper as follows: 
We first introduce the background of BOSM platforms in Section~\ref{sec2}.
Then, we depict the methodology of data collection in Section~\ref{sec3}.
In Section~\ref{sec4}, we analyze downvoting from three aspects, individual downvoting, mutual downvoting and topics o downvoted posts.
Finally, we discuss related work in Section~\ref{sec5} and conclude in Section~\ref{sec6}.

\section{Background}
\label{sec2}
A blockchain is a decentralized distributed ledger that maintains a record of all transactions on the chain\cite{b8}.
Blockchain is decentralized and tamper-resistant because every node in a P2P network has the same copy~\cite{b9}. 
In addition, blockchain also has unique economic properties. 
Therefore, a large number of Blockchain Online Social Media (BOSM) platforms~\cite{b10} emerged recently.
These BOSM platforms address the key issues of traditional OSM platforms, including a single point of failure and lack of incentives, while also being resistant to censorship and promoting the authenticity and truthfulness of content submitted on the platform. 
Next, we will introduce Steemit and a few other active BOSN platforms.

\subsection{Steemit}
Steem is a social blockchain based on the Delegated Proof of Stake (DPoS) consensus protocol. Steemit is a BOSM platform built on the Steem blockchain.

Steemit, unlike other social media platforms that typically do not reward users, offers three categories of incentives: (1) producer rewards, (2) author rewards, and (3) curatorial rewards. 
Steemit replies on a committee of 21 users (called witnesses) to jointly manage the platform and leverage Producer Reward to incentive users to participate in the election of the committee.
To incentivize users to create high-quality content, Steemit periodically allocates Author Reward to users who create posts based on the votes received by these posts.
In addition, to encourage users to vote for posts, Steemit periodically allocates Curatorial Reward to users who vote for posts~\cite{b11}.

On the Steemit platform, users can post content and comment on the content of other users. The content can be voted on by others to demonstrate its quality. Voting by users is separated into upvote and downvote, with upvote indicating positive voting and downvote representing negative voting. The posts with the most votes will have the opportunity to be promoted to the homepage, which will also result in more rewards for the authors and curators. In addition, Steemit includes a reputation system in which users' reputation scores grow as they receive more upvotes and lowers as they receive more downvotes.

\subsection{Other BOSMs}
\textbf{Sapien} is an Ethereum-based social news network that utilizes blockchain technology to produce `social news for the masses.' Content shared on Sapien can be made public or private, ensuring a certain level of social data visibility.

\textbf{Peepeth} is an Ethereum-based online social networking platform similar to Twitter. It consists of two parts: the database composed of Ethereum~\cite{b12} and IPFS~\cite{b13}; and the Peepeth front-end. Actions such as posting, liking and following on Peepeth require payment of Gas fees to package on the chain. To encourage users to create high-quality and authentic content, Peepeth has a once-a-day `Like' function for each user and is irrevocable.

\textbf{Indorse} is a decentralized professional social network based on blockchain technology. Users can earn rewards for sharing their skills and endorsing others' skills multiple times, and advertisers can use IND tokens to buy ad space on the platform. In addition, Indorse seeks to deliver LinkedIn-like features via blockchain technology.

\textbf{Hive Blog} is an online social media based on the Hive blockchain. It is a fork of the Steem blockchain and shares many similarities with Steem.

In addition to these BOSMs, there are a number of additional prominent platforms, including Minds, Appics, etc.

\section{Data collection}
\label{sec3}

\begin{table}
\caption{Schema of operation vote}      
\label{t1}
\small
\begin{center}
\begin{tabular}{|p{2cm}|p{1cm}|p{4.5cm}|}
\hline
{\textbf{Field name}} & {\textbf{Type}} & {\textbf{Description}} 
\\ \hline
    block\_no & Integer & the block recording this operation \\
    \hdashline 
    voter & String & voter's account name \\
    \hdashline 
    author & String & author's account name \\
    \hdashline 
    permlink & String & unique string identifier for the post \\
    \hdashline 
    weight & Integer & weight of vote \\
    \hline
\end{tabular}
\end{center}

\end{table}

\begin{figure*}[htbp]
\centering
\subfigure[2016Q2]{
\includegraphics[width=3.8cm]{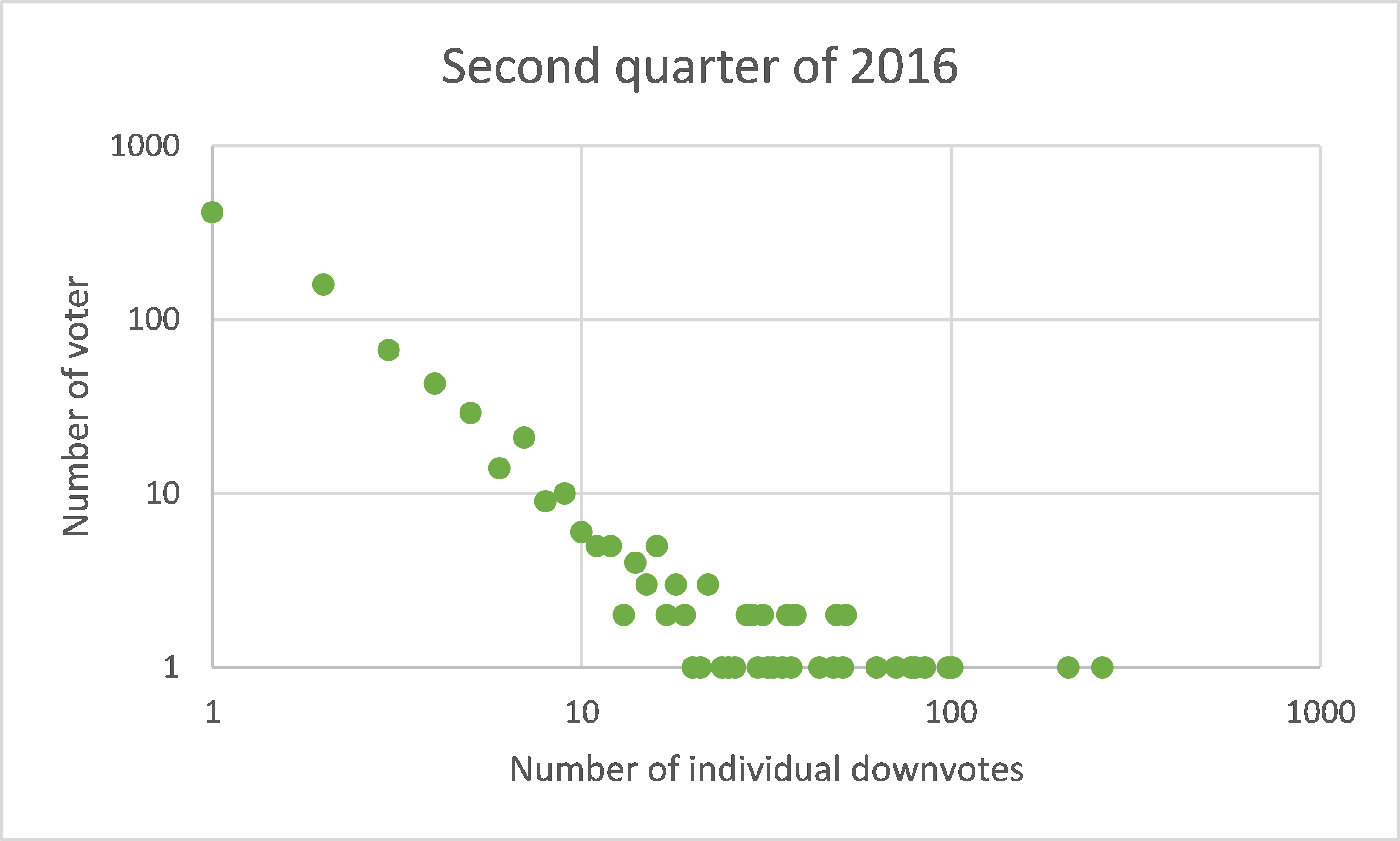}
}
\quad
\subfigure[2016Q3]{
\includegraphics[width=3.8cm]{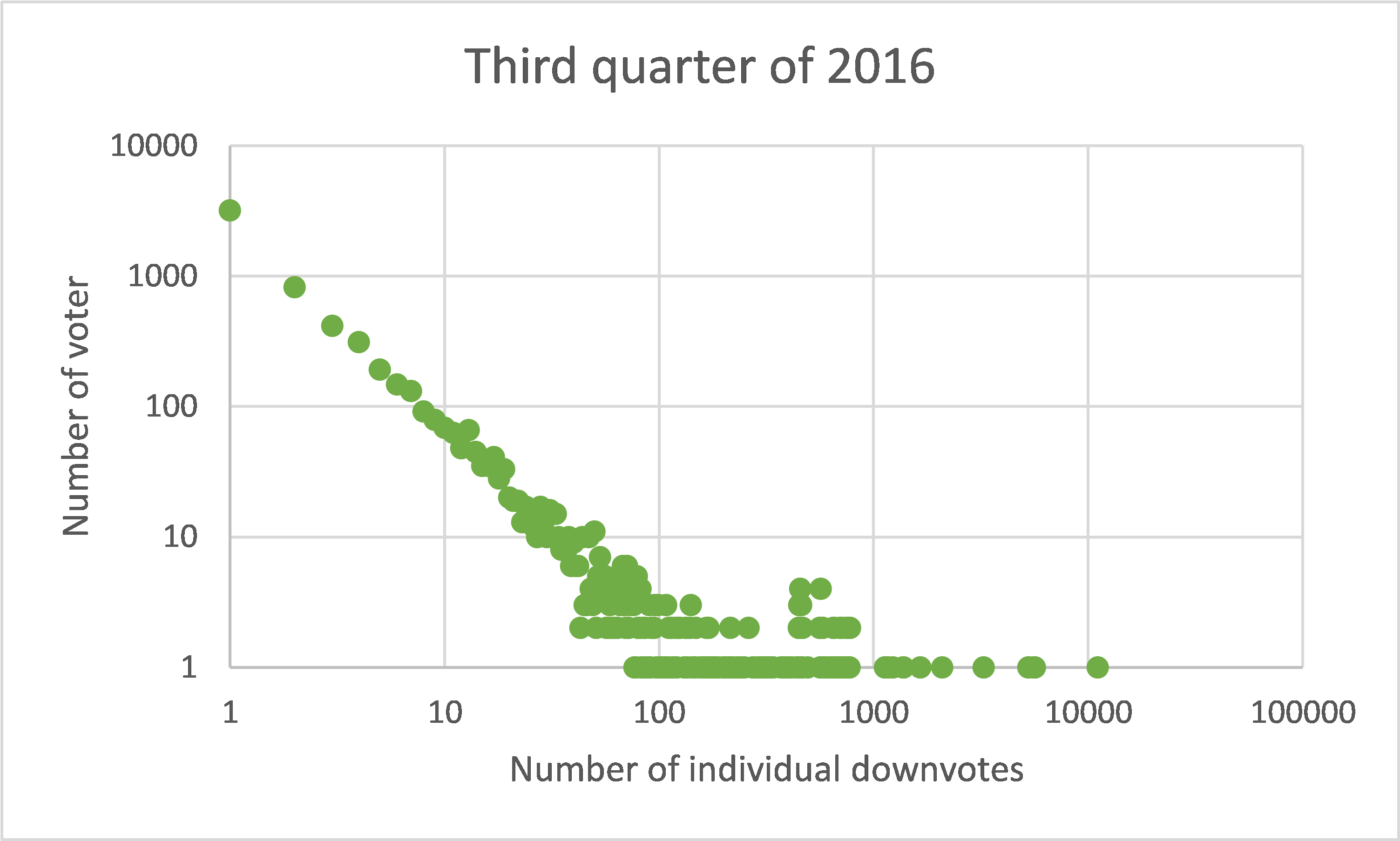}
}
\quad
\subfigure[2016Q4]{
\includegraphics[width=3.8cm]{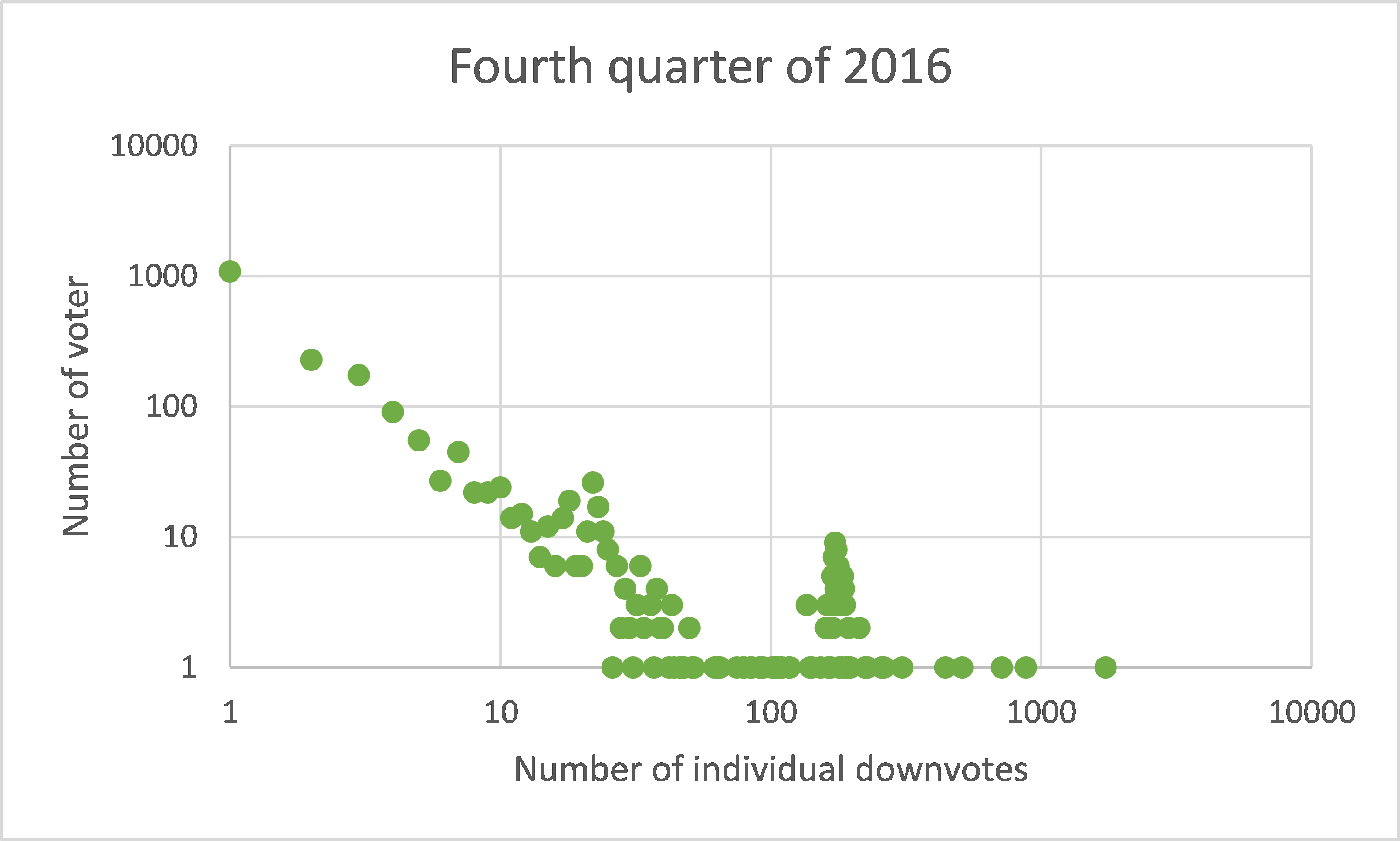}
}
\quad
\subfigure[2017Q1]{
\includegraphics[width=3.8cm]{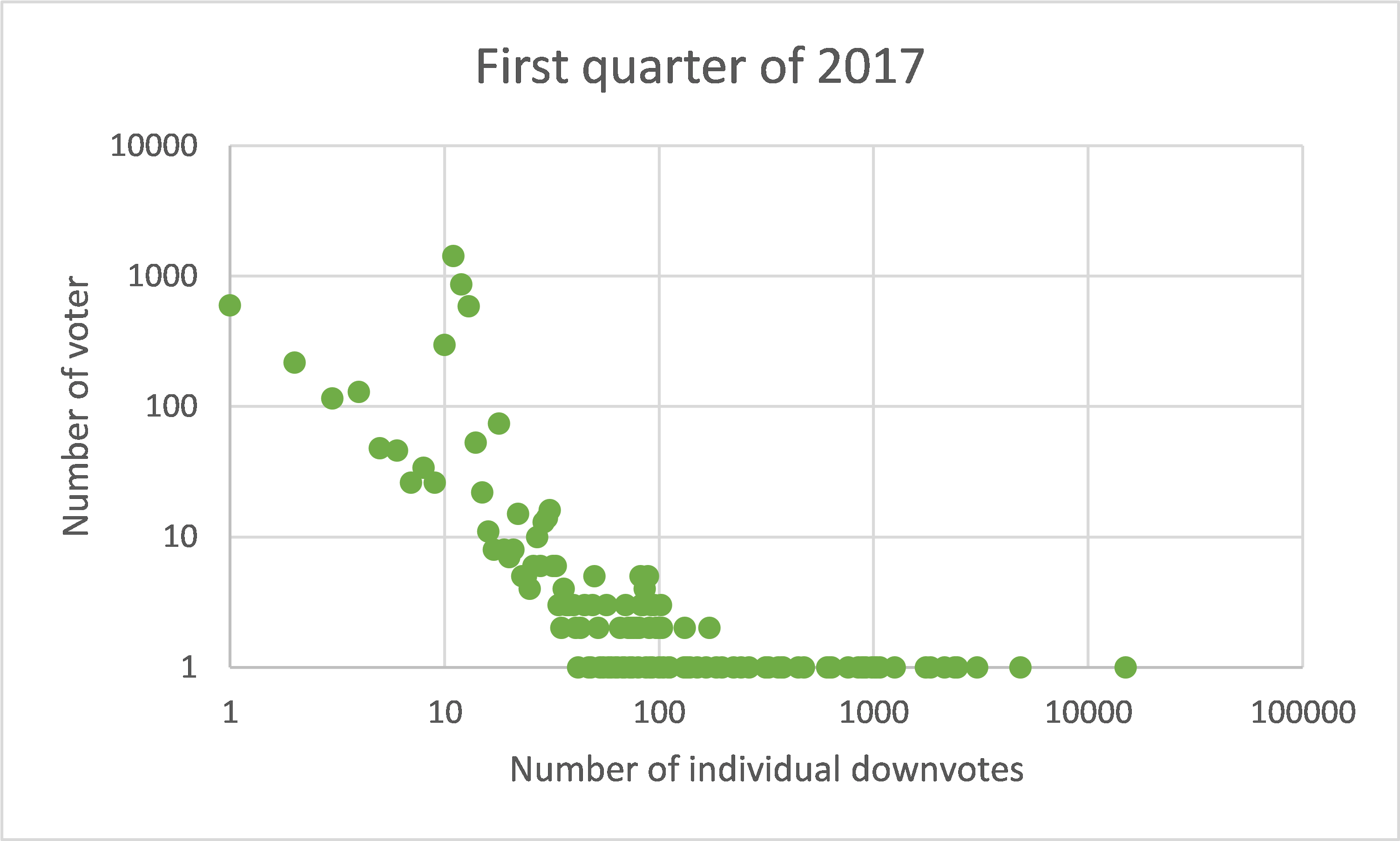}
}
\quad
\subfigure[2017Q2]{
\includegraphics[width=3.8cm]{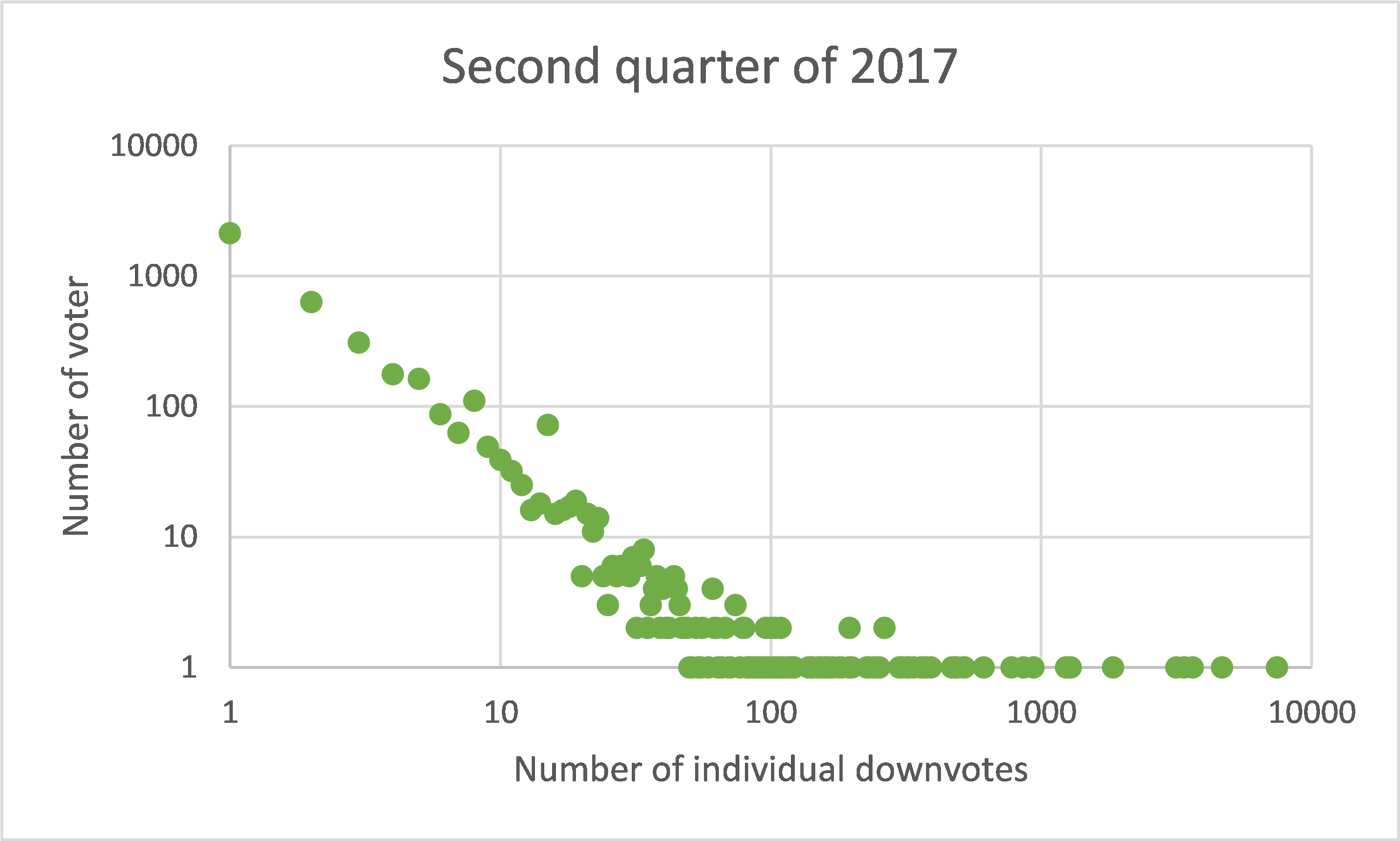}
}
\quad
\subfigure[2017Q3]{
\includegraphics[width=3.8cm]{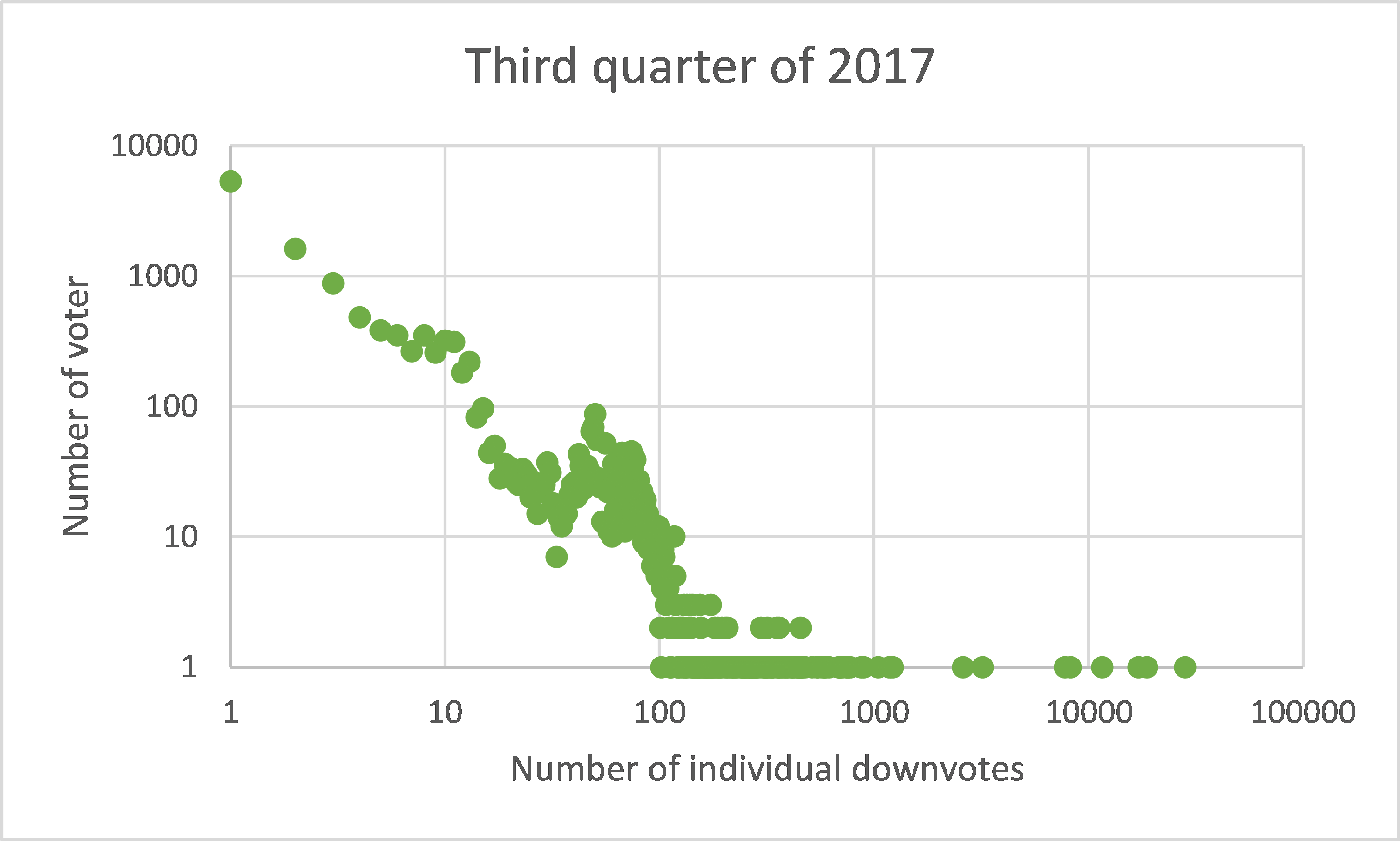}
}
\quad
\subfigure[2017Q4]{
\includegraphics[width=3.8cm]{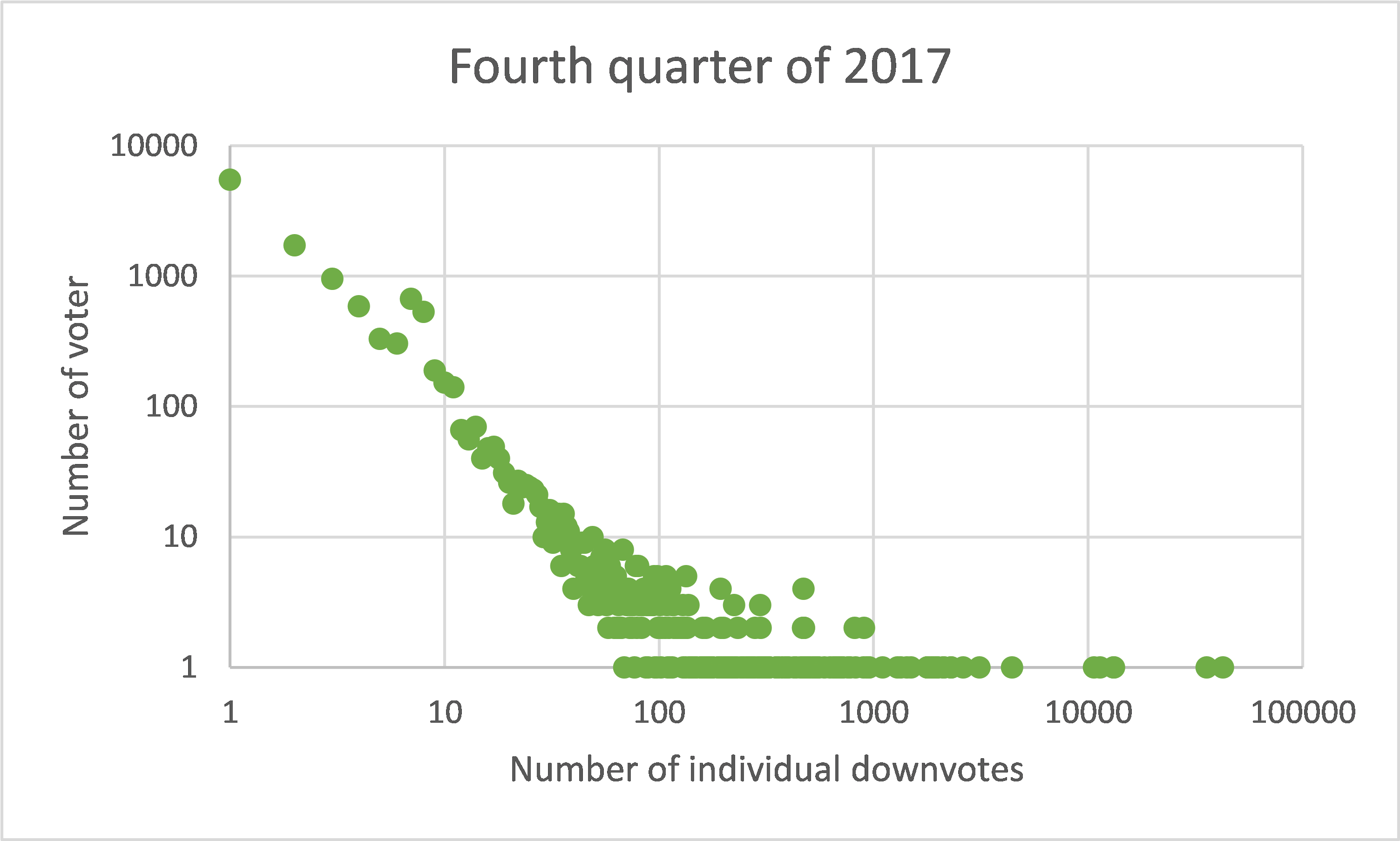}
}
\quad
\subfigure[2018Q1]{
\includegraphics[width=3.8cm]{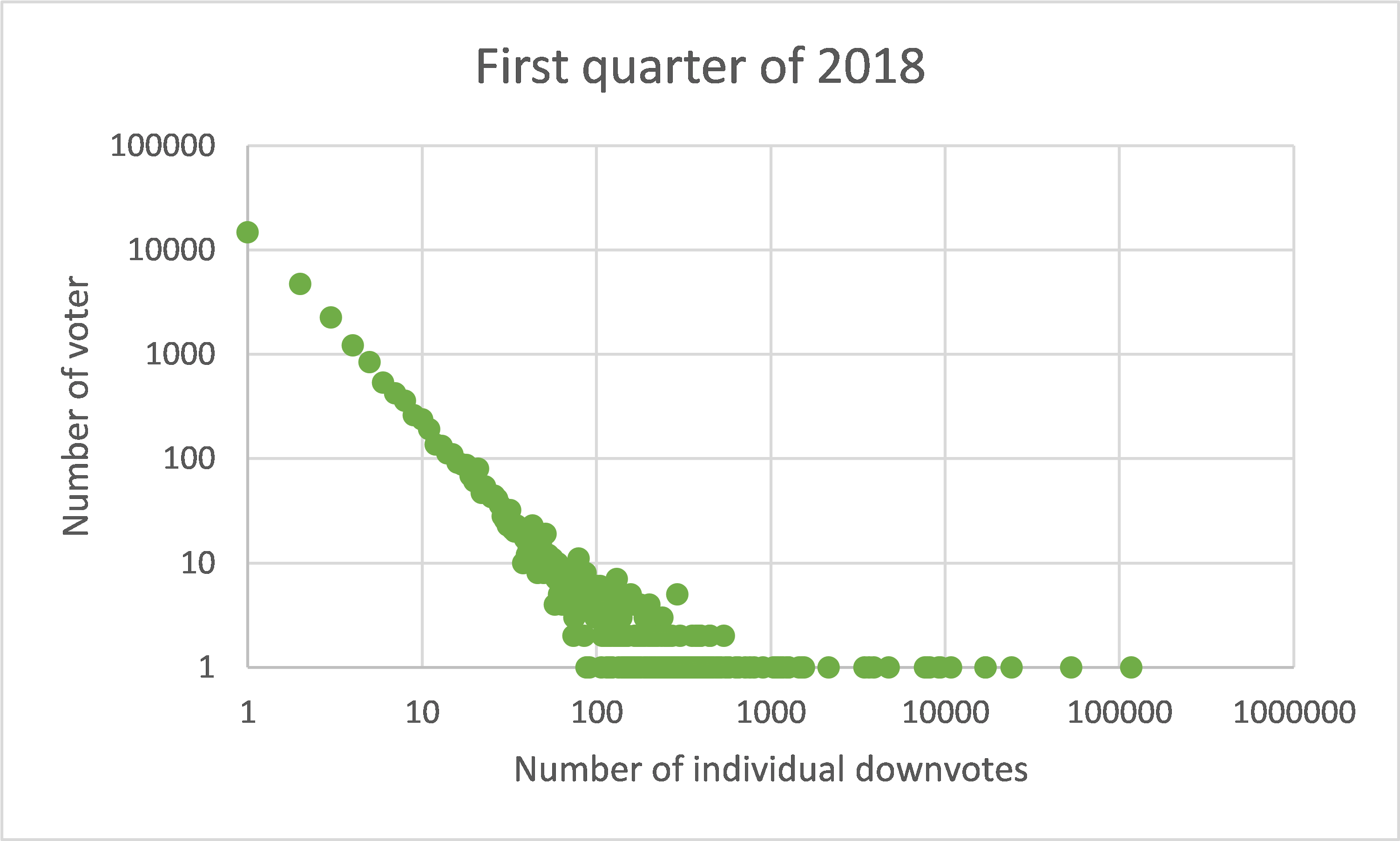}
}
\quad
\subfigure[2018Q2]{
\includegraphics[width=3.8cm]{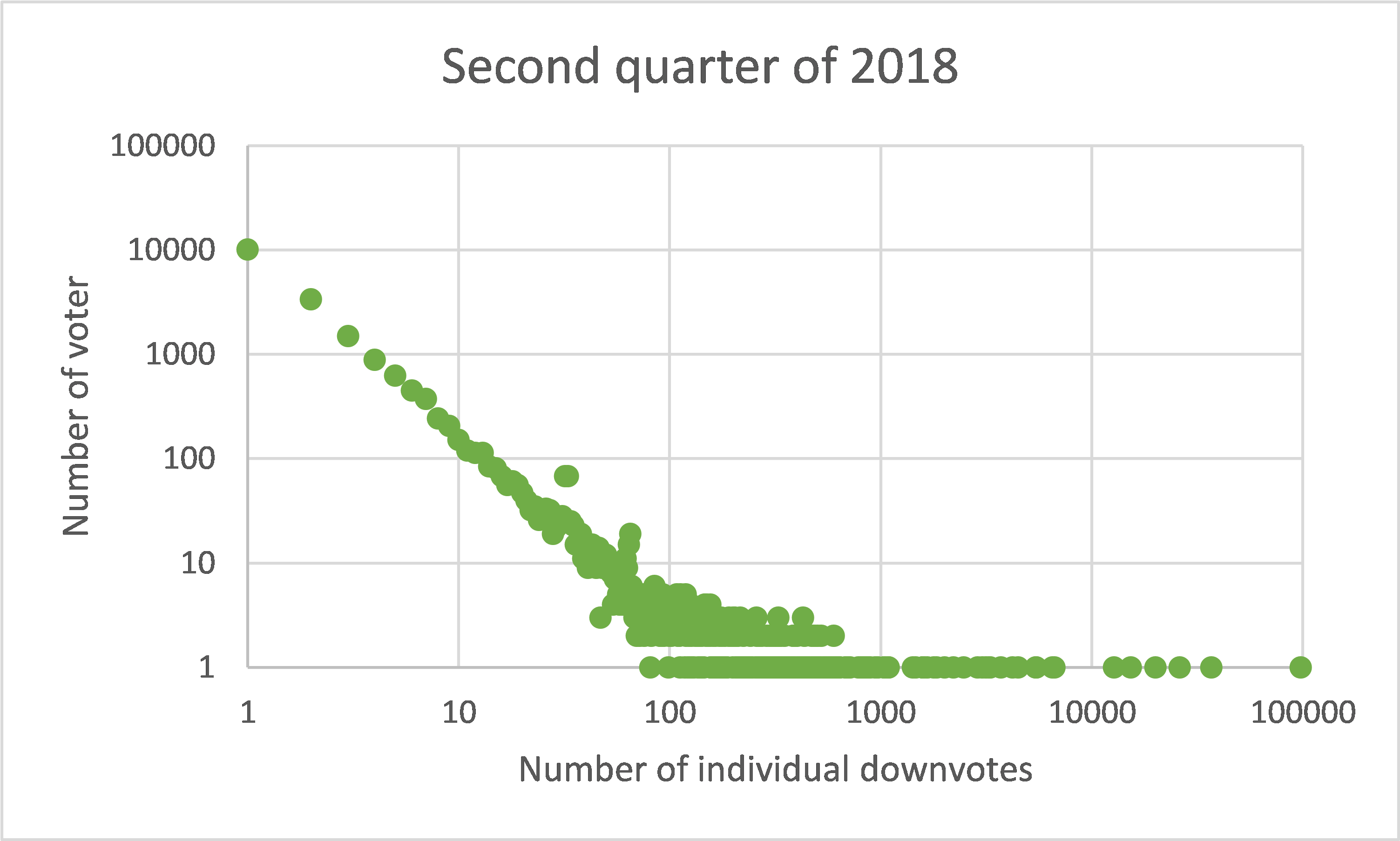}
}
\quad
\subfigure[2018Q3]{
\includegraphics[width=3.8cm]{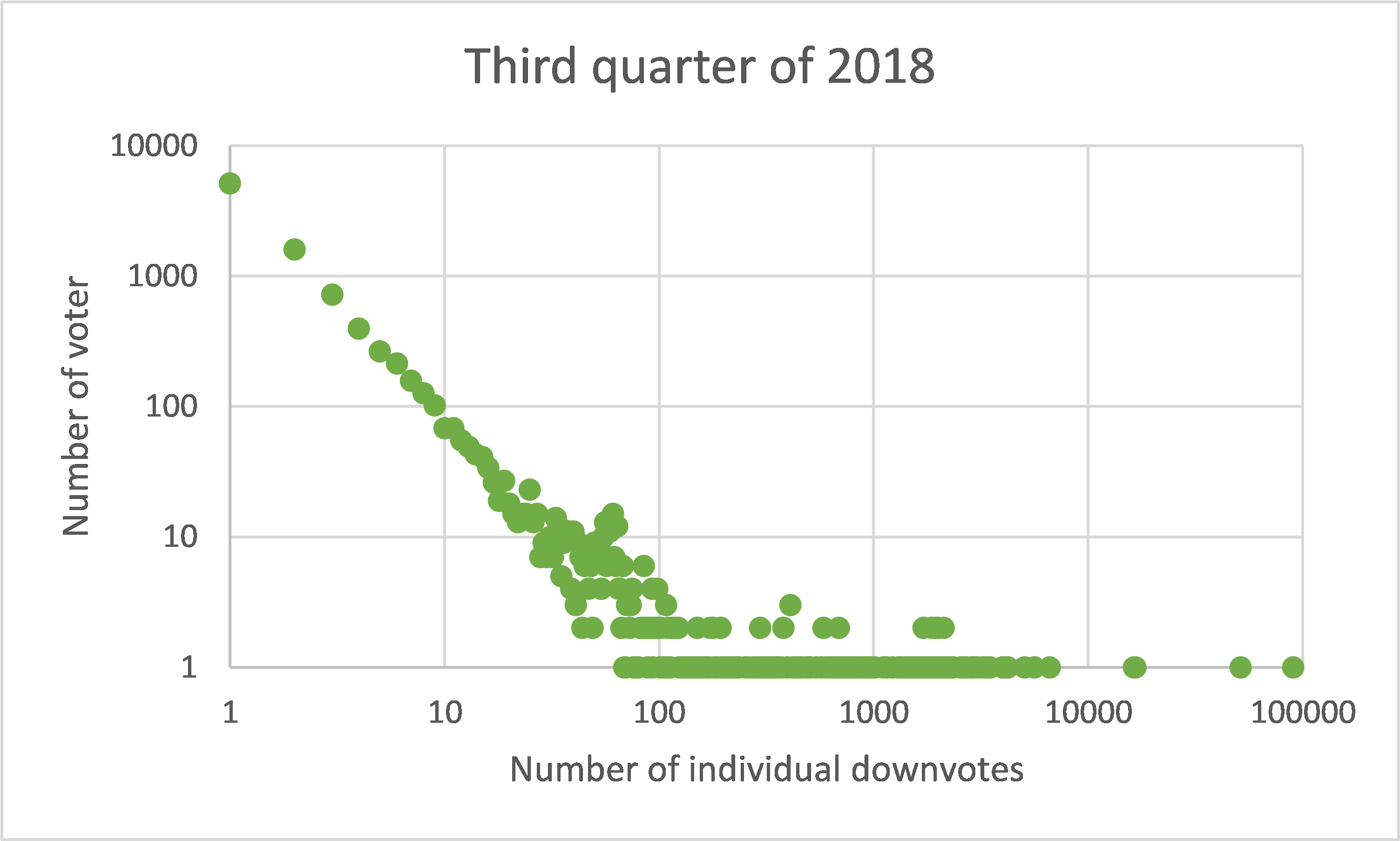}
}
\quad
\subfigure[2018Q4]{
\includegraphics[width=3.8cm]{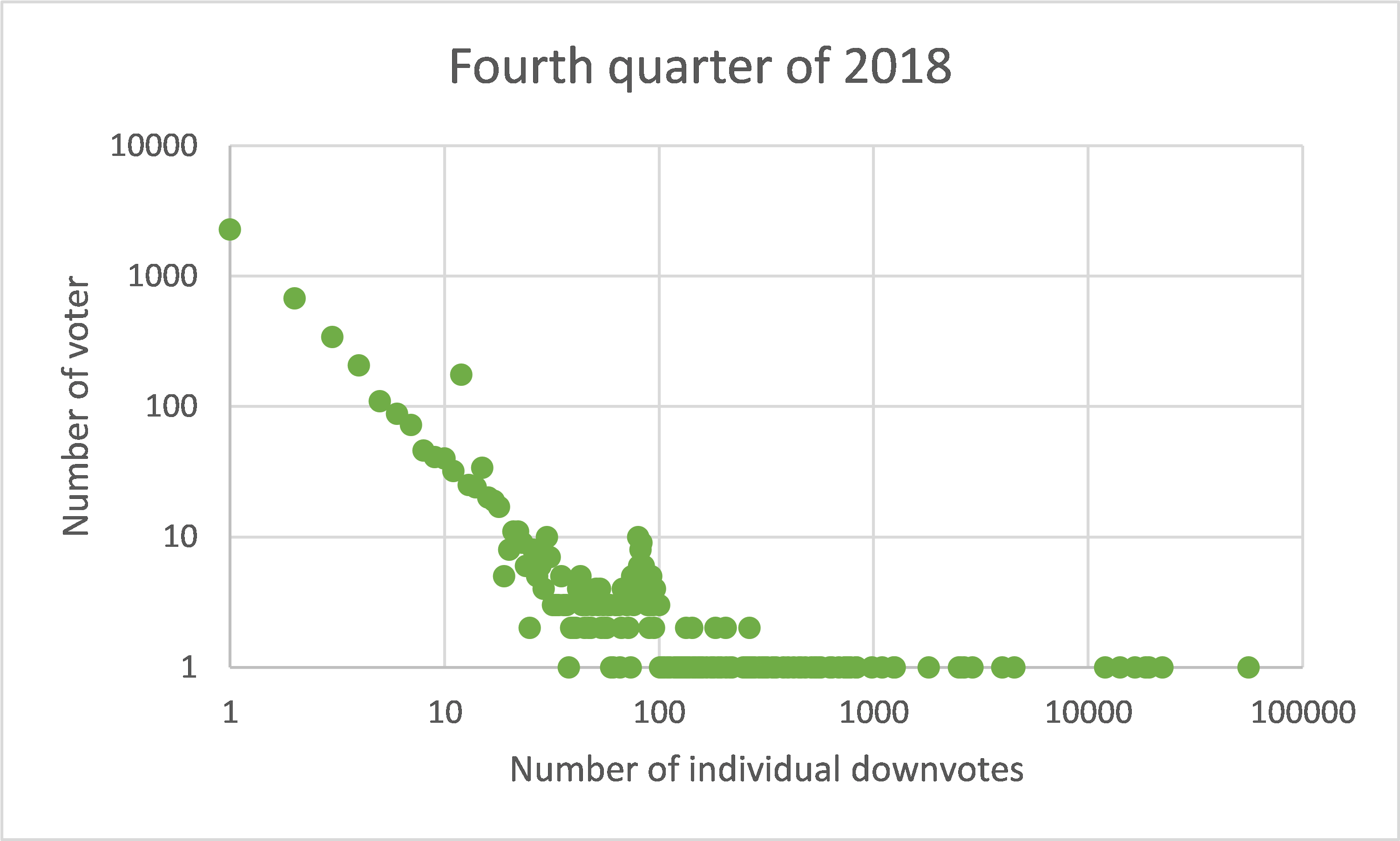}
}
\quad
\subfigure[2019Q1]{
\includegraphics[width=3.8cm]{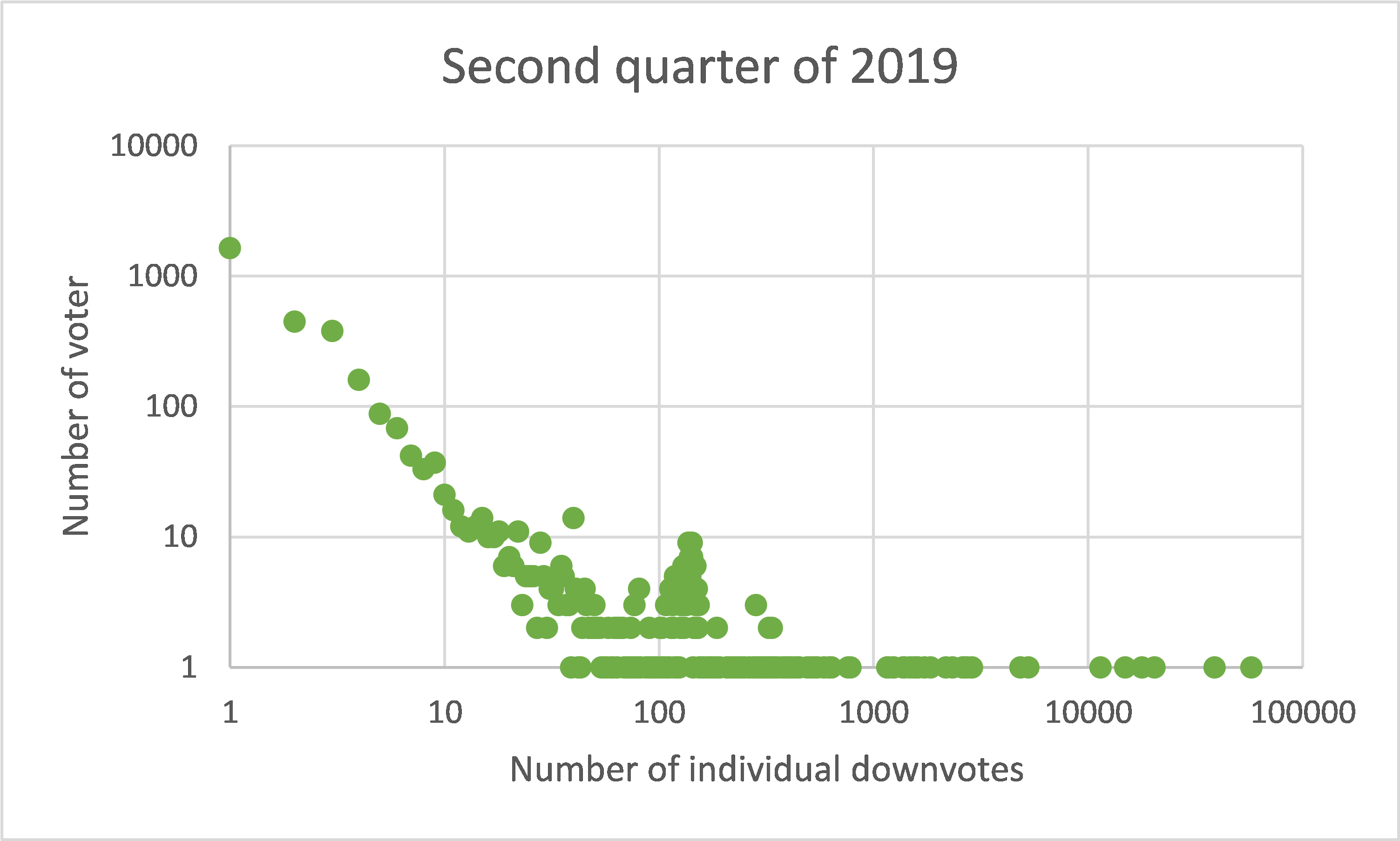}
}
\quad
\subfigure[2019Q2]{
\includegraphics[width=3.8cm]{2019q2.png}
}
\quad
\subfigure[2019Q3]{
\includegraphics[width=3.8cm]{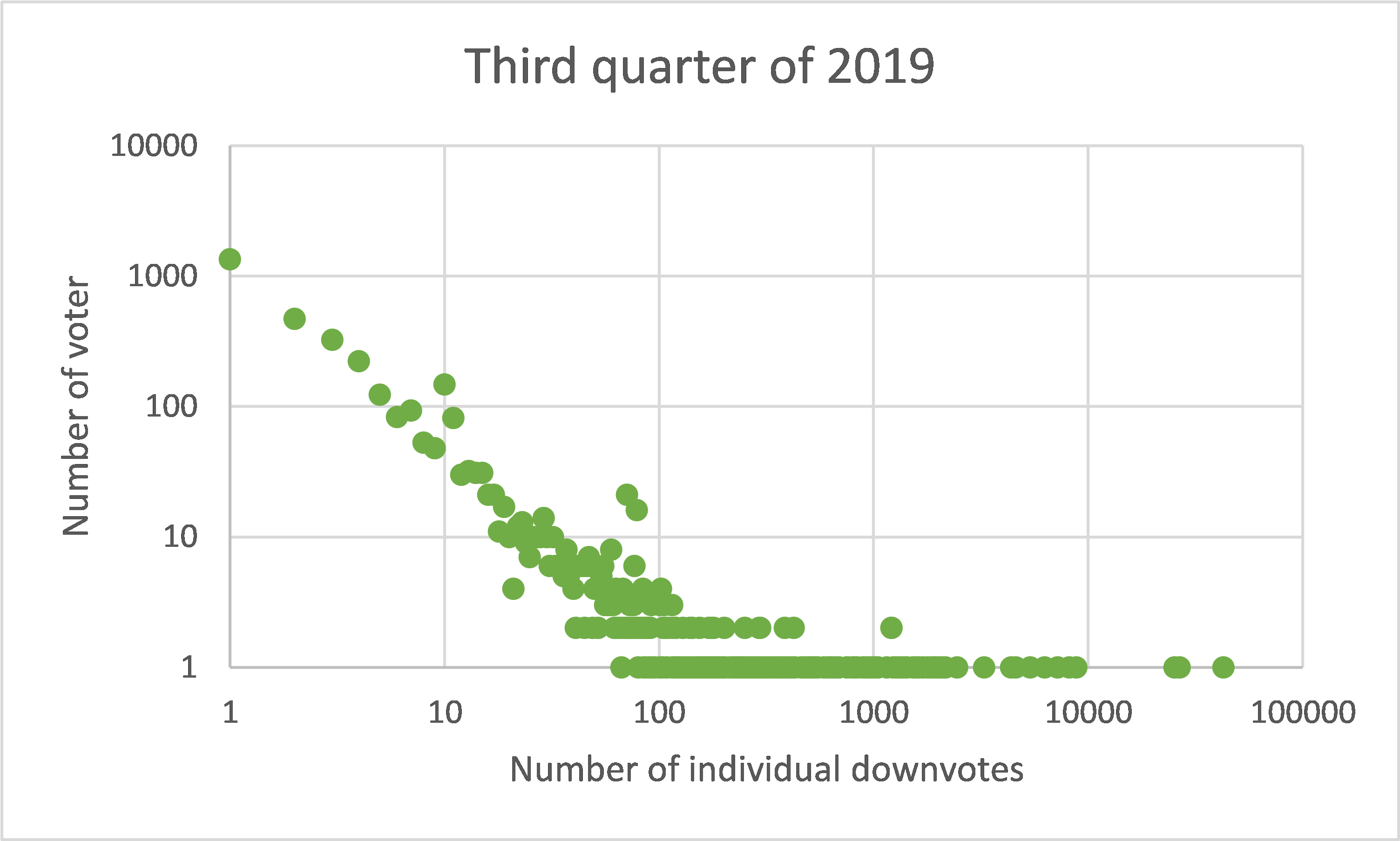}
}
\quad
\subfigure[2019Q4]{
\includegraphics[width=3.8cm]{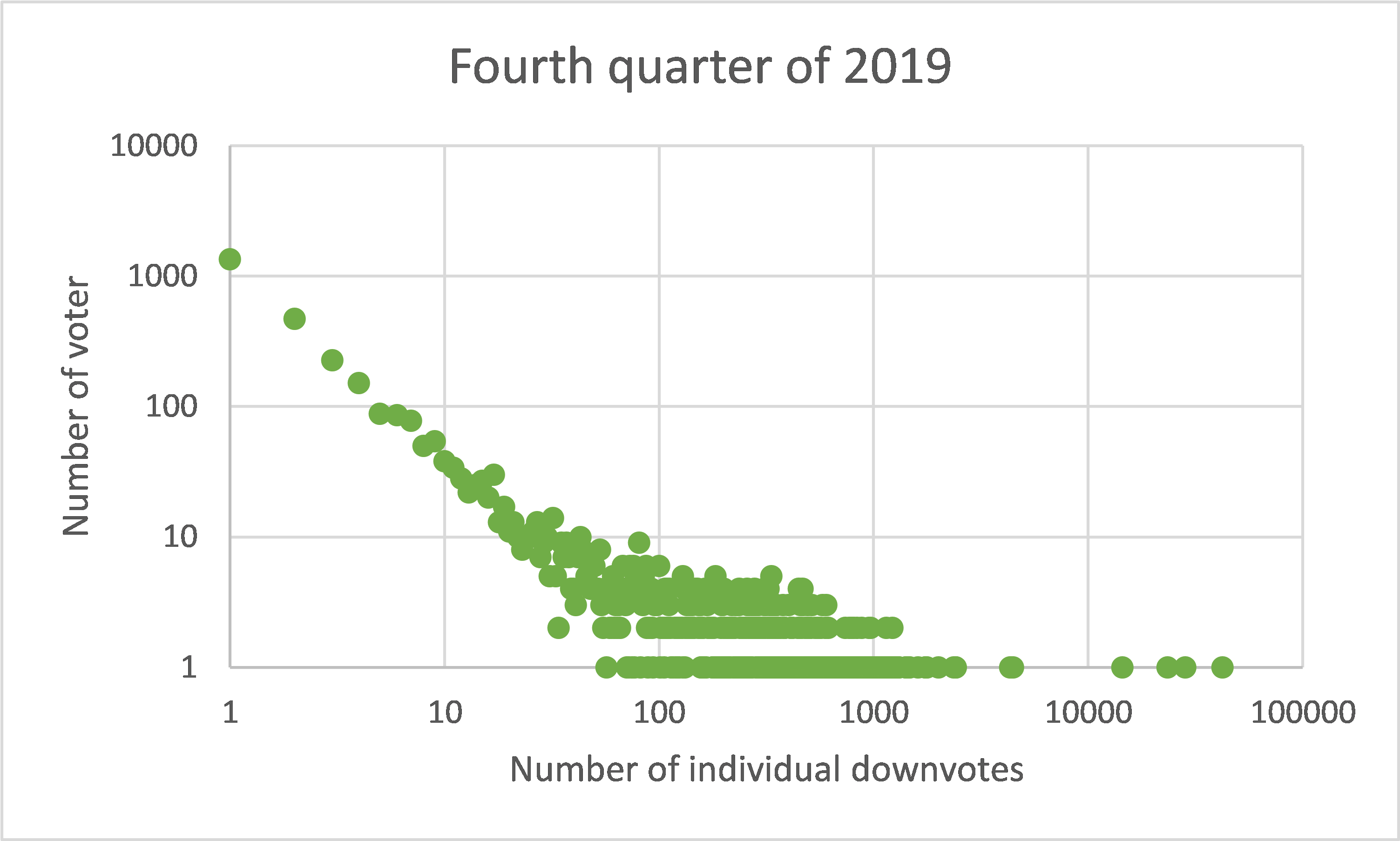}
}
\quad
\subfigure[2020Q1]{
\includegraphics[width=3.8cm]{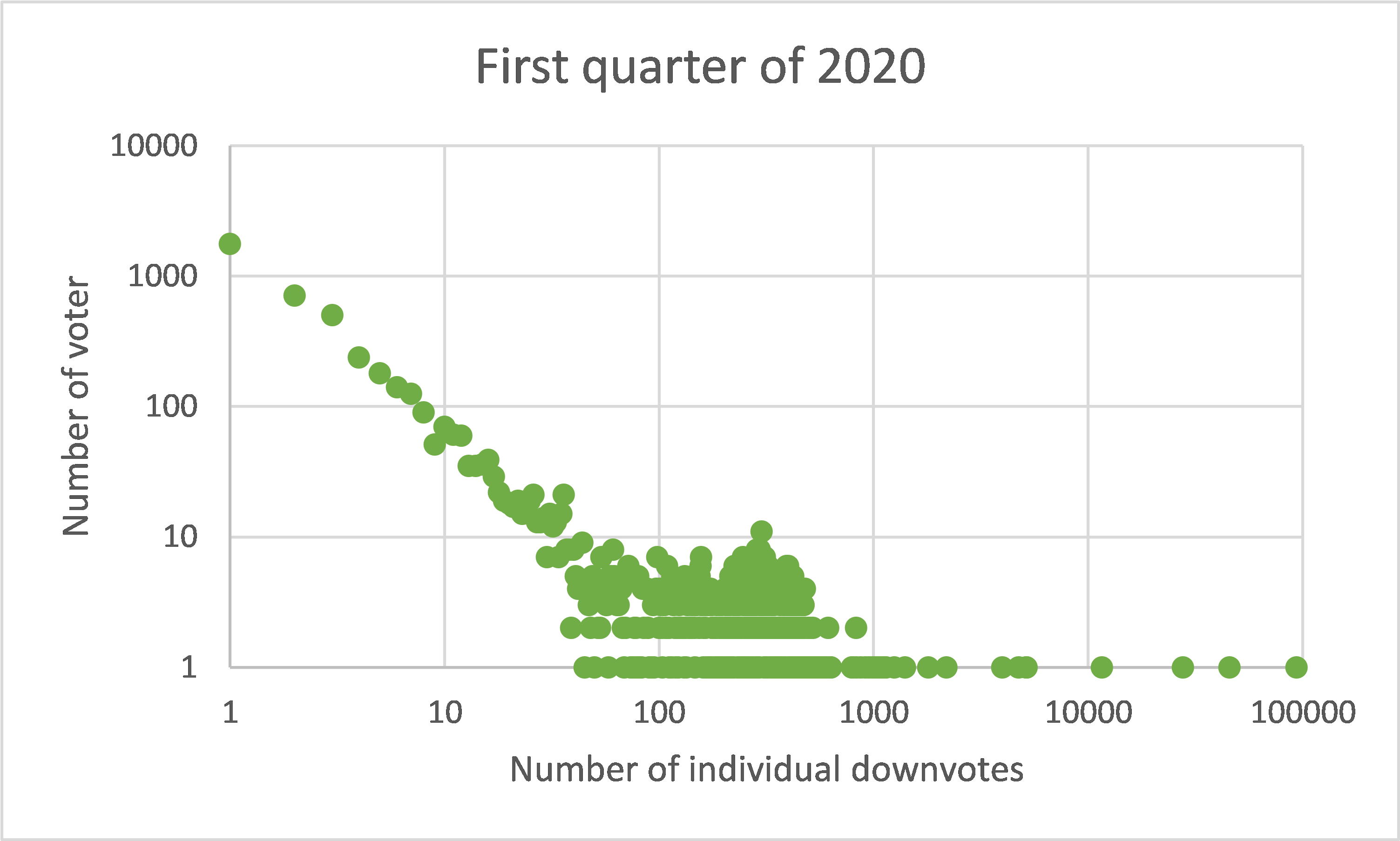}
}

\caption{Statistics of Individual Downvoting}
\label{1}
\end{figure*}

Using the Interactive Application Programming Interface (API) provided by the Steem blockchain, we have gathered transactions from April 2016 to March 2020, a four-year period of time. The obtained data can be categorized as follows: social operation data, witness election operation data, and money transfer operation data~\cite{b14}. 
In our experiments, we utilize primarily the social operation data, including operation \textit{vote} and operation \textit{comment}.
In TABLE~\ref{t1}, we show the schema of operation \textit{vote}, the most important type of operation in our          research.
As can be seen, this operation includes five fields.
It indicates that a person has voted with a specific weight on a post or comment.
A user may set the voting weight $vw$ to any value between -100\% and 100\%.
Specifically, if a user wishes to downvote a post, $vw$ should be adjusted between -100\% and 0\% in order to reduce the accumulated voting power of the downvoted post.
In contrast, to upvote a post, a user should set $vw$ between 0\% and 100\% such that the accumulated voting power gained by the upvoted post is increased.
For instance, a user, say $u_a$, may feel uncomfortable about a post entitled $title$, which was written by another user, say $u_b$. The user $u_a$ may then decide to downvote this post by submitting an operation with information $u_a,u_b,title,-100\%$ to the blockchain.
If this operation is included in a block with No. $bn$, the final operation recorded in the blockchain would be $vote=\{bn,u_a,u_b,title,-100\%\}$, as illustrated in TABLE~\ref{t1}.
Besides, \textit{Steemit} utilizes voting power $vp$ to limit the number of weighted votes cast per day by users.
Each user begins with $vp=100\%$.
If a user continues to vote, then his or her $vp$ will continue to decrease.
Each day, 20\% of $vp$ is recovered.
Intuitively, a user may cast $k$ votes with full voting weight $vw=100\%$ per day, including $k_u$ upvotes and $k_d$ downvotes, where $k=k_u+k_d$.

\section{Analysis of Downvoting}
\label{sec4}

In this section, we study and analyze downvoting in BOSM using our dataset introduced in Section~\ref{sec3}.
Recent research has demonstrated that many social network features can be abused in ways not intended by their designers, thereby causing severe harm to the platform's ecosystem~\cite{b15}.
For instance, authors and curators may purchase upvotes from other users or bot accounts to raise the accumulated voting power of their posts, hence increasing their chances of receiving higher author and curator awards.
However, motivations for downvoting may be very different than those for upvoting.
A voter may earn curatorial rewards for upvoting posts, but there are no explicit benefits for downvoting.
As introduced in Section~\ref{sec3}, the total number of upvotes and downvotes that each user can cast per day can be considered as a fixed value $k$.
Thus, to some extent, downvoting may be considered to be performed at the expense of the personal interests of users.
To understand downvoting, this section focuses on three aspects, namely individual downvoting, mutual downvoting and the topics of downvoted posts.


\begin{figure}
\centerline{\includegraphics[scale=0.12]{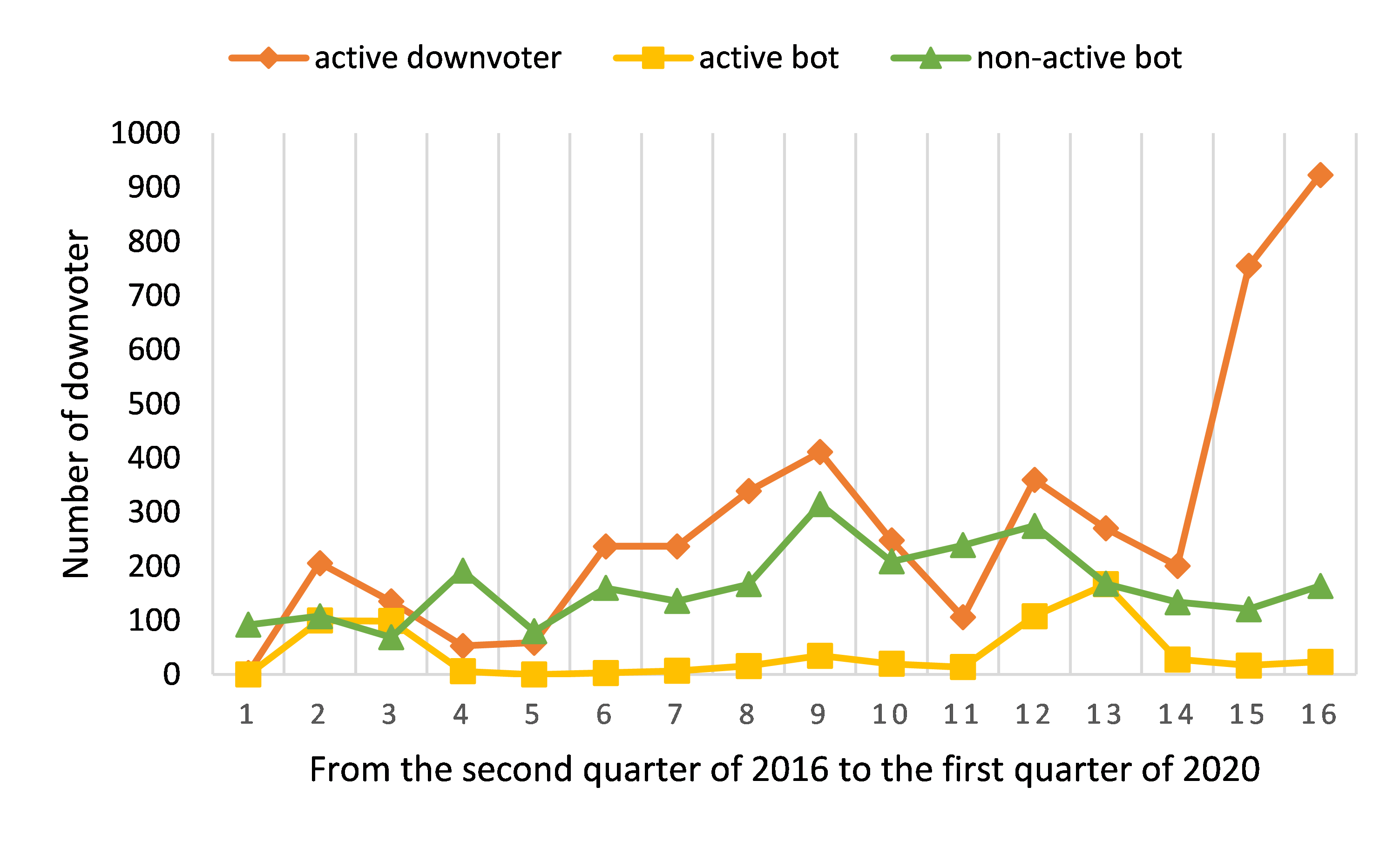}}
\caption{Statistics of the number of active downvoters, active bot users and non-active bot users.}
\label{comp}
\end{figure}


\subsection{Individual Downvoting}
In this section, we extracted and analyzed the number of individual downvotes and voters for each quarter from 2016Q2 to 2020Q1.
The results are shown in Fig.~\ref{1}. 

First, we noted that a large proportion of voters in each quarter cast a single vote, comprising 45.71\% of the total number of voters. 
We noted that the majority of users (82.56\% of the total number of voters) cast no more than 10 votes. 
We consider users with more than 100 votes to be active downvoters who may prefer to vote negatively. According to the statistics, the number of users who have cast more than 100 downvotes, or active downvoters, only accounts for 3.55\% of the total number of users.

After obtaining the statistical results, we focused on the active downvoters. 
Specifically, on the basis of their IDs, we first analyzed these active downvoters manually.
We discovered a large number of similarly named accounts, whose IDs were typically composed of a word and a number, such as `cheetah01'.
Therefore, we defined such accounts as suspected bot users and filtered such bot users whose IDs were of the above-mentioned form. 
We further define the suspected bot users whose IDs contain the same word (e.g., `cheetah'in`cheetah01') as a group of Sybil bots controlled by a single entity.
Then, by observing the votes cast by Sybil bots, we discovered interesting instances in which a group of Sybil bots followed a single leader to downvote a set of posts, as well as instances in which no such leader existed.
Moreover, by comparing active downvoters and bot users, we discovered that only a small percentage of bot users are active downvoters who cast more than 100 votes every month.
Fig.~\ref{comp} presents the statistics of the number of active downvoters, active bot users and non-active bot users.
As can be seen, active bot users account for approximately 14.07\% of all active downvoters, while bot users account for approximately 2.6\% of all downvoters.


Based on our analysis, we can conclude that the majority of users who downvote only do so once. While there were instances where a high number of downvoters also voted for other options, we found that this may have been due to an increase in the number of posts during that time. 
Furthermore, a significant number of downvotes were made by bots. In quarters with fewer bots, the overall trend of downvoting was downward. However, in quarters with a higher bot presence, there were some spikes in downvoting activity. Therefore, in order to maintain a healthy ecosystem, our findings suggest that Steemit should take measures to address the use of bots. For instance, it could consider banning bot users from voting or restricting their access to the platform's activities.

\subsection{Mutual Downvoting}

\begin{figure}
\centerline{\includegraphics[scale=0.12]{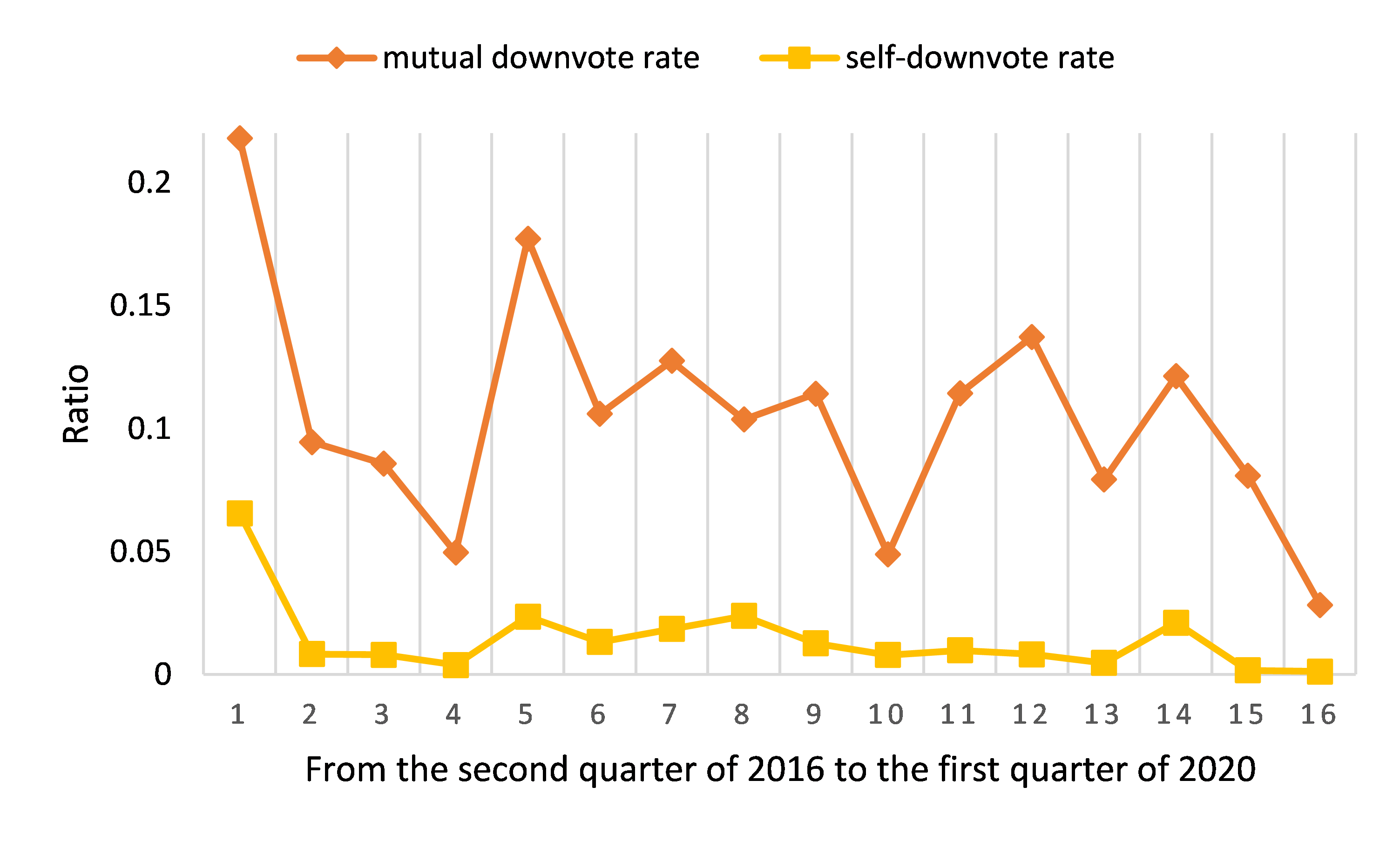}}
\caption{Statistics of mutual downvote rate and self-downvote rate.}
\label{2}
\end{figure}

During our observation of the collected data, we discovered instances of mutual downvoting, namely the behavior that two users downvote each other.
To understand the underlying reasons for this behavior, we conducted an analysis of mutual downvoting. 
Specifically, we sought to understand whether users were engaging in retaliation by downvoting another user's posts, or if they were engaging in other types of mutual voting behavior. 
For example, users may downvote another user's posts only because that user downvoted their posts earlier.
Users may also downvote posts with content similar to their own ones to increase the rewards for their own posts.

We computed the ratio of mutual downvotes (i.e., `mutual-downvote') to total downvotes, as well as the ratio of self downvotes (i.e., `self-downvote') to total downvotes, and showed the results in Fig.~\ref{2}.
In the figure, the red dash indicates the mutual-downvote rate for each quarter, and the blue dash represents the self-downvote rate for each quarter.
Due to the limited sample size, the proportion of mutual downvoting is about 22\% during the first few quarters. 
However, as the sample size increases, the percentage of mutual downvoting tends to decrease to an average of about 9\%.
The trend of the self-downvote rate is comparable to that of mutual downvoting, which trends toward the mean as the sample size increases by roughly 1\%.



After conducting a thorough manual analysis of user self-downvoting and mutual downvoting, we found that many users downvote their own posts to test the system. This is understandable, as nobody wants to suffer losses. We also discovered that mutual downvoting behavior between users did not always follow the expected pattern. This behavior can be influenced by contradictions in each other's replies, downvotes given to each other, and even multiple posts being downvoted. Furthermore, we observed cases where bots were used to downvote the same account multiple times.

\subsection{Topics of Downvoted Posts}
In \cite{b16}, the authors employ Latent Dirichlet Allocation(LDA) to perform natural language processing on the topics of user posts on Steemit to identify the content of interest to Steemit users. 
In order to figure out whether there were targeted downvotes for certain content, we also examined the topics of the posts that received downvotes and compared them to the overall topic distribution.
LDA is a topic model. It is a three-layer Bayesian probability model that takes word, topic, and document structure into account~\cite{b17}. The model can be presented in the form of a probability distribution of the document's topic, and LDA can be used to detect whether a document or corpus has latent topic information.

We choose LDA to extract the topics of all the downvoted posts. 
We manually set the number of categories to 6 and output the words corresponding to the topics, as shown in Table~\ref{tab2} below, we did not filter the numerical data during the data pre-processing, so there are numerical words in the categories.

\begin{table}[htbp]
\caption{The Topics of Downvoted post}
\begin{center}

\begin{tabular}{|c|c|c|}
\hline
Number & Topics & Words \\
\hline
Topic 0 & Time & 2018,2017,2019,2020,2016 \\
\hline
Topic 1 & Money & Steem, free, blockchain, bitcoin, money \\
\hline
Topic 2 & Post & News, animal, themarkymark, price, reward \\
\hline
Topic 3 & Steemit & Steemit, photography, bot, abusereports, power \\
\hline
Topic 4 & Content & Market, contest, challenge, upvote, steamsteem \\
\hline
Topic 5 & Crypto- & Actifit, Cryptocurrency, crypto, bitcoin, coin \\
\hline

\end{tabular}

\label{tab2}
\end{center}
\end{table}

\begin{figure}
\centerline{\includegraphics[scale=0.12]{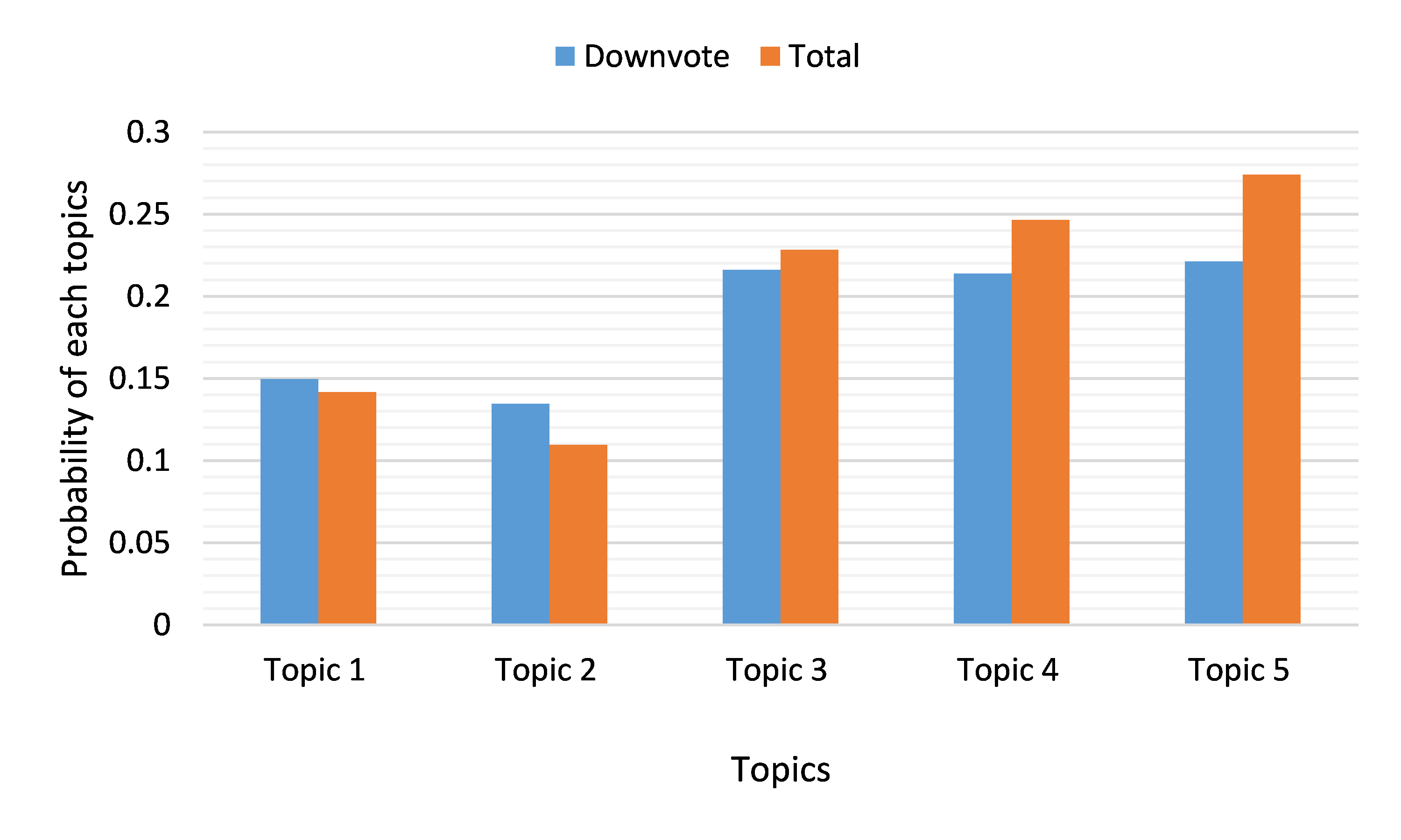}}
\caption{Comparison of the Topic Probability of DownvoteD Posts with the Topic Probability of all Posts.}
\label{fig}
\end{figure}

In Figure~\ref{fig}, the probability of the topic of the downvoted posts measured in this paper is compared to the probability of the topic of all the posts measured in~\cite{b16},
Since there is no topic of years in \cite{b16}, the data of our Topic 0 was removed from the comparison. 
Although there is a slight difference between the downvoted post and the overall distribution of topics, the overall distribution of topics is comparable.
And, given that we have a small margin of error in estimating the probability of \cite{b16}, our results may suggest that there is no targeted downvote for a specific topic.

\section{Related Work}
\label{sec5}
Over the past few years, many studies have studied BOSM platforms from multiple perspectives.
In \cite{b18}, it is stated that bot users are generally less active than human users. Also, we are unable to screen bots by reputation due to their generally high reputation. Also, in \cite{b19}, automated detection techniques for bots and fraudulent activity were developed, and thousands of bot accounts (over 30\% of accounts on the platform) and some real-world attacks (301 attacked accounts) were discovered.
A recent study~\cite{b4} of the economic factors of Steemit revealed that the wealthiest users on the platform do not obtain wealth by being the most active users on the platform, but rather by purchasing cryptocurrency via external mechanisms. The study also assessed the impact of the underlying cryptocurrency on platform user behavior. When the price of Steem is low, people are less encouraged to post content and comments, according to the findings. Nevertheless, user engagement has no significant impact on the price of Steem. In addition, the research demonstrated that somewhat successful users have a greater understanding of ways to obtain more rewards, as opposed to having produced better content.

In addition to identifying anomalous social operation behavior in Steemit, the activity of users in the DPoS consensus process is also examined in recent works. In~\cite{b20}, the gradual evolution of a decentralized system into a monopoly is discovered, along with anomalous voting gangs with similar voting behavior. In~\cite{b21}, authors analyzed the degree of decentralization in both Bitcoin and Steem. Compared to Steem, Bitcoin tends to be more decentralized among top miners but less decentralized in general, as indicated by their research.

Previous studies have either studied user roles, analyzed user behavior from an economic perspective, and some have analyzed user behavior from the perspective of STEEM consensus mechanisms. This paper is based on existing research on bot users, and by analyzing user downvote behavior, we combined internal and external factors that affected user behavior, so as to further identify the abnormal behaviors that existed in users and the reasons for these behaviors.

\section{Conclusion}
\label{sec6}
This paper presented the first empirical study of downvoting on BOSM platforms.
We evaluated individual downvoting behaviors, examined downvoting behaviors between users, and identified and analyzed topics from posts that were downvoted.
Our main findings included a significant number of suspected bot accounts that were actively downvoting content, around 9\% of downvoting that might be retaliatory, and there are no notable instances of content downvoting for a particular topic.
Our findings suggested that downvoting on BOSM platforms may have been abused by a small number of users, but it has not been abused on a large scale.
We believed that the findings in this paper will facilitate the future development of user behavior analysis and incentive pattern design in both BOSM and Web3.




\end{document}